\documentclass[aps,prl,twocolumn,superscriptaddress,preprintnumbers,floatfix,nofootinbib,notitlepage,showkeys]{revtex4-1}

\usepackage[utf8]{inputenc}
\usepackage{cancel}
\usepackage{float}
\usepackage{graphicx}
\usepackage{hyperref}
\usepackage{latexsym}
\usepackage{amsmath}
\usepackage{amssymb}
\usepackage{bbm}

\usepackage{ulem}
\usepackage{pdfsync}
\usepackage{epsfig}
\usepackage{epstopdf}
\usepackage{subfigure}
\usepackage{color}
\usepackage{comment}
\usepackage{slashed}

\def\beq{\begin{equation}}
\def\eeq{\end{equation}}
\def\baq{\begin{eqnarray}}
\def\eaq{\end{eqnarray}}

\newcommand{\bea}{\begin{eqnarray}} 
\newcommand{\eea}{\end{eqnarray}}

\newcommand{\bmp}{\noindent\begin{minipage}{16cm}}
\newcommand{\emp}{\end{minipage}\vskip 7mm} 
\def\lsim{\mathrel{\raise.3ex\hbox{$<$\kern-.75em\lower1ex\hbox{$\sim$}}}}
\def\gsim{\mathrel{\raise.3ex\hbox{$>$\kern-.75em\lower1ex\hbox{$\sim$}}}}

\newcommand{\intron}[1]{}

\begin{document}
\title{Dark matter from scalar field fluctuations}

\author{Tommi Tenkanen}
\email{ttenkan1@jhu.edu}
\affiliation{Department of Physics and Astronomy, Johns Hopkins University, \\
Baltimore, MD 21218, USA}

\begin{abstract}
Dark matter (DM) may have its origin in a pre-Big Bang epoch, the cosmic inflation. Here, we consider for the first time a broad class of scenarios where a massive free scalar field unavoidably reaches an equilibrium between its classical and quantum dynamics in a characteristic time scale during inflation and sources the DM density. The study gives the abundance and perturbation spectrum of any DM component sourced by the scalar field. We show that this class of scenarios generically predicts enhanced structure formation, allowing one to test models where DM interacts with matter only gravitationally.
\end{abstract}

%
\maketitle

%

Dark matter (DM) may have its origin in a pre-big-bang epoch. It may have been produced, for example, by decays or annihilations of particles during the Big Bang, i.e. by the so-called 'freeze-in' \cite{McDonald:2001vt,Hall:2009bx,Bernal:2017kxu} mechanism, or by e.g. the misalignment mechanism which generated a non-zero DM abundance during cosmic inflation (see e.g. Ref. \cite{Marsh:2015xka}). In all such cases, it is crucial to assess not only if enough DM was produced but also that perturbations in DM energy density overlap with those in radiation at large scales to a high precision, i.e. that the so-called DM isocurvature perturbations are within the stringent limits obtained from the Cosmic Microwave Background radiation (CMB) \cite{Akrami:2018odb}.

In this Letter, we study the requirements for fulfilling the above criteria in scenarios where the DM resides in a hidden sector which also contains a scalar field. As scalar fields are typically abundant in extensions of the Standard Model \cite{Arvanitaki:2009fg,Marsh:2015xka,Stott:2017hvl}, their dynamics during inflation is expected to provide the generic initial conditions for non-thermal production of DM after inflation. Another possibility is that the scalar field(s) themselves constitute all or part of the observed DM abundance.

Here, we consider for the first time a class of scenarios where a scalar field acting as or sourcing the DM is a massive free field which attained an equilibrium between the classical drift and stochastic quantum fluctuations during inflation. We also make a detailed comparison with the case where the dynamics of the field was determined by the misalignment mechanism and, in both cases, pay particular attention to isocurvature perturbations, evaluating the conditions under which the cosmological constraints on them are avoided.

Similar ideas have recently been studied in the literature, including e.g. self-interacting \cite{Peebles:1999fz,Kainulainen:2016vzv,Heikinheimo:2016yds,Enqvist:2017kzh,Markkanen:2018gcw} or non-minimally coupled DM \cite{Cosme:2017cxk,Cosme:2018nly,Alonso-Alvarez:2018tus}, DM coupled to the inflaton \cite{Bertolami:2016ywc}, or axion DM \cite{Beltran:2006sq,Graham:2018jyp,Guth:2018hsa,Ho:2019ayl}. In this Letter we show for the first time that even in the simplest possible case the scalar field can successfully constitute all DM without being in conflict with the CMB data. In particular, we show that the scenario does not require a specific initial misalignment but under suitable conditions the scalar field can start at the minimum (or, in fact, at any value) and it will reach an equilibrium state which determines the final DM abundance. We generalize our results by discussing also scenarios where the scalar field sources only a part of the DM density and evaluate the resulting perturbation spectrum also in that case. Finally, we present a novel finding that this class of scenarios generically predicts enhanced structure formation, allowing one to test models where DM interacts with matter only gravitationally. We also discuss how to rule out the present model.


As a benchmark scenario, we consider the simplest possible DM Lagrangian
\begin{equation}
\mathcal{L}_{\rm DM}= \frac12\partial^\mu\chi\partial_\mu\chi - \frac12m^2\chi^2 ,
\end{equation}
where $\chi$ is a scalar field. We assume that $\chi$ is decoupled from radiation and minimally coupled to gravity\footnote{To make a comparison with previous studies in the literature \cite{Cosme:2017cxk,Cosme:2018nly,Alonso-Alvarez:2018tus}, we note that in case of a non-minimal coupling to gravity of the type $\xi\chi^2 R$, where the Ricci scalar during inflation is proportional to the Hubble scale as $R=12H_*^2$, the study in this Letter corresponds to the case $12\xi \ll m^2/H_*^2$.}. We assume standard cosmological history and that inflation was driven by something other than $\chi$, which we assume was energetically subdominant during inflation. We assume the field responsible for inflation was also responsible for generating the initial curvature perturbation and reheating the Universe after inflation.

Assuming the field $\chi$ was light during inflation, $m/H_* < 1$, it was displaced from its low energy minimum and 
gained a non-zero expectation value $ \chi_*^2 \equiv \langle \chi^2 \rangle$ in our observable Hubble patch (see e.g. Ref. \cite{Riotto:2002yw}). Therefore, at the end of inflation there was an effective scalar condensate with a non-zero energy density, which together with the corresponding fluctuation spectrum provides the initial conditions for post-inflationary dynamics. However, as we will show, not all values of $\chi_*$ are preferred nor accepted by inflationary dynamics and cosmological constraints. 

The equation of motion for the field describing its post-inflationary dynamics is
\begin{equation}
\label{eom}
\ddot{\chi} + 3H\dot{\chi} + m^2\chi = 0,
\end{equation}
which is solved for
\begin{equation}
\label{chi_solution}
\chi(t) = 2^{1/4}\Gamma\left(\frac{5}{4}\right)\chi_*\frac{J_{1/4}(mt)}{\left(mt\right)^{1/4}} ,
\end{equation}
where $J_\nu$ is the Bessel function of rank $\nu$. Throughout this Letter, we assume that the Universe was radiation-dominated from the end of inflation, so that $H=1/(2t)$. At late times, the solution \eqref{chi_solution} oscillates rapidly with an amplitude
\begin{equation}
\chi_0(t) = \frac{2\Gamma\left(\frac{5}{4}\right)}{\sqrt{\pi}}\frac{\chi_*}{(mt)^{3/4}}\,, \hspace{.5cm} mt \gg 1,
\end{equation}
and the field has associated energy density
\begin{equation}
\label{rho_chi}
\rho_\chi = \frac12m^2\chi_0^2 \simeq \frac{\Gamma^2\left(\frac{5}{4}\right)}{\pi}\frac{\sqrt{m}\chi_*^2H_*^{3/2}}{a^3} ,
\end{equation}
where we have fixed the scale factor $a_*=1$ at the end of inflation. The result agrees very well with the approximation that the field was initially frozen at $\chi_*$ and the oscillations started when the field became massive at $m \simeq 1.5 H_{\rm osc}$.

As shown by Eq. \eqref{rho_chi}, at late times the field $\chi$ acts effectively as a cold dark matter component. Its contribution to the total DM abundance at the present day is thus given by
\begin{equation}
\label{DMabundance}
\frac{\Omega_\chi h^2}{0.12} = 3.5\times 10^{17}g_*^{-1/4}(H_{\rm osc})\left(\frac{\chi_*}{M_{\rm P}}\right)^2\sqrt{\frac{m}{{\rm GeV}}} ,
\end{equation}
where $M_{\rm P}$ is the reduced Planck mass and $g_*(H_{\rm osc})$ is the effective number of entropy degrees of freedom at $H_{\rm osc}$, for which we will use the Standard Model value $g_*(T\gg 100\, {\rm GeV})=106.75$ for simplicity. Thus, for suitable values of $\chi_*$ and $m$, the field can constitute all of the observed DM. 


Because the field is assumed to be decoupled from radiation, fluctuations in the local scalar field value necessarily generate isocurvature perturbations between the DM and radiation energy densities. The isocurvature perturbation is defined in the usual way as
\begin{equation}
\label{isocurvature_def}
S_{r\chi} \equiv 3H\left(\frac{\delta\rho_r}{\dot{\rho_r}} - \frac{\delta\rho_\chi}{\dot{\rho_\chi}} \right),
\end{equation}
where perturbations are defined as deviations from the background, $\delta\rho_i \equiv \rho_i/ \langle\rho_i\rangle -1$. Because the fluids are assumed to be decoupled from each other, we obtain $-H\delta\rho_\chi/\dot{\rho_\chi} = \delta\rho_i/(3(1+w_i)\rho_i) \equiv \delta_i/(3(1+w_i))$, where $\delta_i$ is the density contrast of the fluid $i=r,\chi$ and $w_i \equiv p_i/\rho_i$ is the equation of state parameter which relates the pressure of the fluid to its energy density. For the scalar field $p_\chi = (\dot{\chi}^2-m^2\chi^2)/2$ and $\rho_\chi = (\dot{\chi}^2+m^2\chi^2)/2$. Because the fluids are decoupled, isocurvature is conserved on super-horizon scales, $\dot{S}_{r\chi}=0$ \cite{Wands:2000dp}.

To highlight the differences between the usual treatment and the novel results presented below, we will first consider a scenario where the field dynamics during inflation is assumed to be dominated by slow-roll. This case corresponds to the usual misalignment mechanism. If $m<H_*$, the field acquired small fluctuations around its mean value $\chi_*$. Assuming that the potential and the associated density perturbation can be expanded linearly in the field, the isocurvature perturbation becomes \cite{Langlois:2004nn}
\begin{equation}
S_{r\chi}  = -\delta_\chi = -2\frac{\delta\chi_*}{\chi_*} ,
\end{equation}
which gives the primordial isocurvature power spectrum as \cite{Riotto:2002yw}
\begin{equation}
\mathcal{P}_S = \left(\frac{2}{\chi_*}\right)^2\mathcal{P}_{\delta\chi} \simeq \left(\frac{2}{\chi_*}\right)^2\left(\frac{H}{2\pi}\right)^2\left(\frac{k}{aH}\right)^{2\eta_\chi - 2\epsilon} ,
\end{equation}
where the result is valid to first order in $3\eta_\chi \equiv m^2/H^2$ and $\epsilon = -\dot{H}/H^2$ is the usual slow-roll parameter characterizing change in the expansion rate during inflation. We assume $\epsilon \ll 1$, which is the case for e.g. a large class of plateau models which are in perfect agreement with the most recent Planck data \cite{Akrami:2018odb}. Therefore, in all the cases we consider, we take $H=H_*$ during inflation. Thus, the isocurvature power spectrum becomes
\begin{equation}
\label{P_SR}
\mathcal{P}_S \approx \mathcal{A}_{\rm sr}\left(\frac{k}{k_*}\right)^{n_\delta - 1} ,
\end{equation}
where
\begin{eqnarray}
\label{AnClass}
\mathcal{A}_{\rm sr} &=&\frac{1}{\pi^2}\left(\frac{H_*}{\chi_*}\right)^2e^{-N_*(n_\delta - 1)} , \\ \nonumber
n_\delta &=& 1+ \frac23\frac{m^2}{H^2} ,
\end{eqnarray}
and
\begin{equation}
\label{efolds}
N_* \simeq 59 + \frac12\ln\left(\frac{H_*}{8\times 10^{13}\,{\rm GeV}}\right)
\end{equation}
is the number of e-folds from the horizon exit of the pivot scale $k_* = 0.002\,{\rm Mpc}^{-1}$ and the end of inflation. From Eq. \eqref{AnClass} we see that the fluctuation spectrum can be strongly blue-tilted, $n_\delta > 1$, for large $m$. 

\begin{figure}[t]
	\begin{center}	\includegraphics[width=0.5\textwidth]{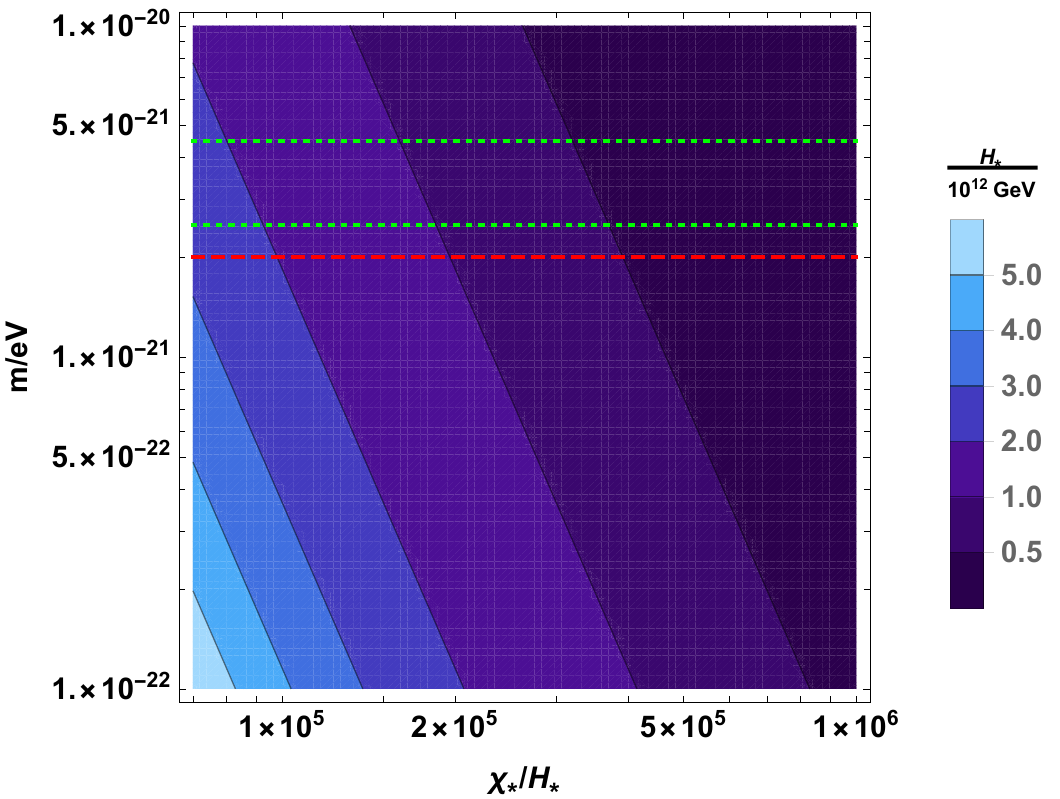}
	\end{center}
	\caption{The parameter space relevant for the 'fuzzy DM' scenario \cite{Hui:2016ltb}. Contours show the value of $H_*$ required for $\chi$ to constitute all DM. The regions below the horizontal red dashed line and between the dotted green lines are in tension with observations of the Lyman-$\alpha$ forest \cite{Irsic:2017yje} and the recent Event Horizon Telescope data due to the black hole superradiance mechanism \cite{Davoudiasl:2019nlo}, respectively. See also Ref. \cite{Marsh:2018zyw} which placed tentative constraints excluding the entire region shown here. The lower limit for x-axis, $\chi_*/H_*\simeq 7\times 10^4$, is given by the isocurvature constraint \eqref{chi/H}.}
	\label{parameters_fuzzy}
\end{figure}

The Planck satellite mission placed stringent constraints on DM isocurvature \cite{Akrami:2018odb}. In particular, for the uncorrelated DM isocurvature considered in this Letter, non-observation of primordial isocurvature in the CMB places a constraint 
\begin{equation}
\label{isocurvature_constraint}
{P}_S\lesssim \beta\mathcal{P}_\mathcal{R}(k_*) ,
\end{equation}
where $\mathcal{P}_\mathcal{R}(k_*)\simeq 2.1\times10^{-9}$ is the observed amplitude of the curvature power spectrum and $\beta \leq 0.011$ \cite{Akrami:2018odb}. We find that for $m^2/H_*^2\ll 1$, the present model avoids the isocurvature constraint for the initial field value
\begin{equation}
\label{chi/H}
\frac{\chi_*}{H_*} \gtrsim \frac{1}{\sqrt{\pi^2\beta\mathcal{P}_\mathcal{R}(k_*)}} \simeq 7\times 10^4 .
\end{equation}
Note that during and also long after inflation the energy density of $\chi$ was still subdominant to the total energy density for small enough $m$. An example of the model parameter space where the scalar field simultaneously constitutes all DM and evades the isocurvature bound \eqref{isocurvature_constraint} is shown in Fig. \ref{parameters_fuzzy} for varying $H_*$. The shown parameter space is relevant for the 'fuzzy DM' scenario \cite{Hui:2016ltb}. For non-negligible $m$ the perturbation amplitude \eqref{AnClass} is even smaller at the pivot scale due to the blue tilt, and the correct DM abundance is obtained for a large variety of masses and initial field values. 

However, for large $m/H_*$ the field will quickly reach a regime where the classical drift responsible for slow-roll and the stochastic quantum fluctuations are in equilibrium. Following the method in Ref. \cite{Starobinsky:1994bd} (see also Refs. \cite{Bertolami:2016ywc,Guth:2018hsa,Markkanen:2019kpv}), one can show that for a quadratic potential the distribution of field values in patches the size of the horizon at the end of inflation is Gaussian with zero mean and a variance given by
\begin{equation}
\label{chi_variance}
\langle \chi^2 \rangle = \frac{3H_*^4}{8\pi^2m^2} .
\end{equation}
Therefore, in this case there is no mean field over the observable Universe but a large distribution of values described by the equilibrium distribution. The equilibrium is naturally attained in $N\simeq H_*^2/m^2$ e-folds regardless of the initial field value \cite{Enqvist:2012xn}. Thus, in this case $\chi_*$ is not a free parameter and, consequently, we will use the typical field value $\chi_* = \sqrt{\langle \chi^2 \rangle}$ when evaluating the DM energy density \eqref{DMabundance}.

Fluctuations in the scalar field energy density are in this case characterized by the power spectrum \cite{Markkanen:2019kpv}
\begin{equation}
\label{stochastic_P}
\mathcal{P}_{\delta_\chi} = \mathcal{A}_{\rm sto}\left(\frac{k}{k_*}\right)^{2(n_\delta - 1)} , 
\end{equation}
where
\begin{equation}
\label{An}
\mathcal{A}_{\rm sto} \approx 4(n_\delta - 1)e^{-2N_*(n_\delta - 1)} ,
\end{equation}
and $n_\delta - 1$ and $N_*$ are given by Eqs. \eqref{AnClass} and \eqref{efolds}, respectively. Because also in this case\footnote{This can be shown in e.g. the synchronous gauge where there are perturbations only in $\chi_*$ and $a$. Note that because in the stochastic case the mean field value vanishes, $\langle \chi\rangle = 0$, it would, in general, be incorrect to assume $\delta_\chi \propto \delta\chi_*/\chi_*$. Instead, the density contrast power spectrum can be computed using the spectral expansion method, first developed in Ref. \cite{Starobinsky:1994bd} and recently studied in detail in Ref. \cite{Markkanen:2019kpv}.} $S=-\delta_\chi$, we obtain $\mathcal{P}_S=\mathcal{P}_{\delta_\chi}$ and can again straightforwardly apply the constraint \eqref{isocurvature_constraint}.

\begin{figure}[t]
	\begin{center}	\includegraphics[width=0.38\textwidth]{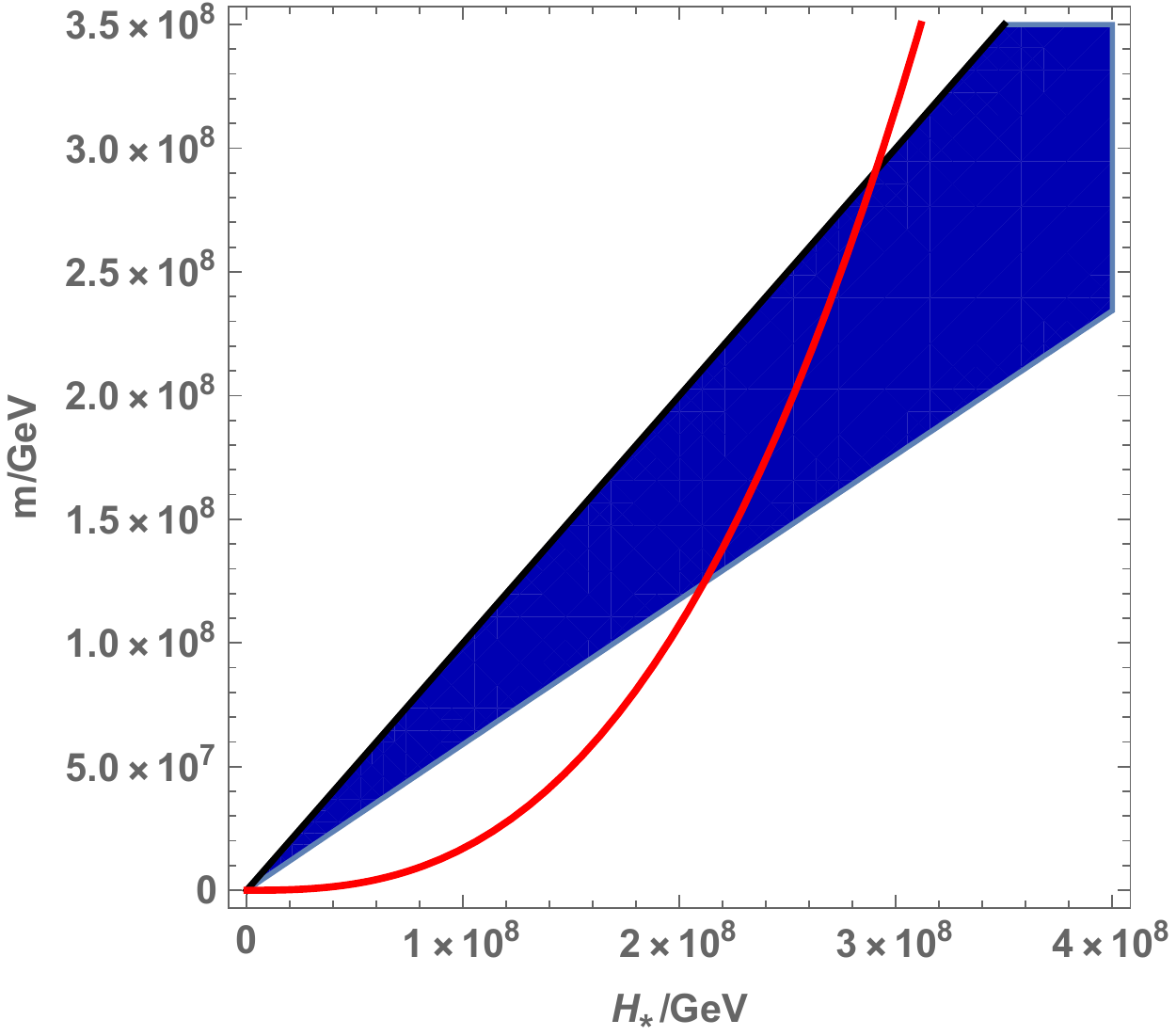}
	\end{center}
	\caption{The region of the model parameter space where the scalar simultaneously constitutes all DM (along the red curve) and avoids the DM isocurvature constraints (blue region). Also the region $m\leq 0.4$ GeV, $H_*\lesssim 10^{5}$ GeV in the lower left corner is allowed. The result is not sensitive to the exact value of $\beta$. Above the blue region, $m>H_*$, the scalar fluctuations during inflation are strongly suppressed.}
	\label{parameters_stochastic}
\end{figure}

Fig. \ref{parameters_stochastic} shows the region of the model parameter space where the scenario simultaneously explains all DM (along the red curve) and avoids the DM isocurvature constraints (blue region). We see that the observed DM abundance is obtained for $H_*\sim m \sim 2\times 10^8$ GeV or $m/H_* < \sqrt{3\beta\mathcal{P}_\mathcal{R}(k_*)/8} \simeq 3\times 10^{-6}$. In the former case the requirement for the initial field value is only $\chi_* \simeq 0.1H_*$, in contrast to $\chi_* \gtrsim 7\times 10^4H_*$ in the classical slow-roll/misalignment scenario, whereas the energy scale of inflation is fixed to $V^{1/4}=(3H_*^2M_{\rm P}^2)^{1/4}\simeq 3\times 10^{13}$ GeV, which in the case of single-field inflation corresponds to a small value of the tensor-to-scalar ratio, $r\sim 10^{-12}$. While such a value is beyond the reach of any foreseeable cosmological mission, it does not require elaborate modifications to the simplest inflationary models compatible with data \cite{Rasanen:2017ivk,Enckell:2018hmo,Antoniadis:2018ywb}. In the latter case above, the constraints are satisfied for any $m\leq 0.4$ GeV when $H_*\lesssim 10^{5}$ GeV (corresponding to $V^{1/4}\lesssim 10^{12}$ GeV). Note that for fixed cosmological parameters, namely $H_*$, this DM model contains only one parameter, $m$, as in the stochastic case the initial value $\chi_*$ is not a free parameter but determined by the equilibrium distribution. Therefore, this scenario constitutes the simplest possible DM model without being sensitive to initial conditions\footnote{The possibility that we live in a patch where the local $\langle \chi^2 \rangle$ differs significantly from the result \eqref{chi_variance} has been discussed in the case of scalar self-interactions in Ref. \cite{Heikinheimo:2016yds}.}.

The analysis conducted above can be easily modified to accommodate also other cosmological histories or models where the DM sector has a richer structure. For example, if the scalar field decayed after inflation into relativistic hidden sector particles $\psi$ which never entered into thermal equilibrium with radiation, their contribution to the present DM abundance is given by
\begin{equation}
\frac{\Omega_\psi h^2}{0.12} = 1.2\times 10^{9}g_*^{-1/4}(H_{\rm osc})\left(\frac{m}{\Gamma_\chi}\right)^{3/8}\left(\frac{\chi_*}{M_{\rm P}}\right)^{3/2}\left(\frac{m_\psi}{\rm GeV}\right) ,
\end{equation}
where $\Gamma_\chi$ is the decay width of $\chi$ and we assumed the $\psi$ particles thermalized with themselves immediately upon the decay of $\chi$ at $H_{\rm dec}=\Gamma$, attaining a temperature $T_\psi^4(H_{\rm dec}) = (30/\pi^2)\rho_\chi(H_{\rm dec})$, and became non-relativistic at $T_\psi = m_\psi$. If the $\psi$ particles were non-relativistic from the beginning, the final DM abundance is given by Eq. \eqref{DMabundance}. In both cases, the perturbation spectrum is given either by Eq. \eqref{P_SR} or \eqref{stochastic_P}, depending on whether the field $\chi$ attained equilibrium between the classical drift and quantum fluctuations during inflation or not.

Other production mechanisms for DM, such as freeze-in, provide an additional contribution to the yield studied in this Letter and the final DM abundance is essentially a sum of the individual contributions \cite{Belanger:2018mqt}. In such cases, the final DM isocurvature perturbation depends on the perturbation spectra of the sources and the ratio of their contributions as \cite{Kainulainen:2016vzv}
\begin{equation}
\label{Src}
S_{rc}  = \frac{\rho_{c}^\chi}{\rho_c^\chi + \rho_c^r}S_{r\chi} ,
\end{equation}
where $\rho_c^i$ is the part of cold DM sourced by the fluid $i=\chi,r$, and $S_{r\chi}$ is defined as in Eq. \eqref{isocurvature_def}. Note that Eq. \eqref{Src} applies independently of the scalar potential.

Finally, we make an important remark about structure formation. Because in the above scenarios the DM field $\chi$ is a genuine isocurvature component, one can expect enhanced structure formation especially at small scales. The curvature perturbation on the uniform total energy density hypersurface is given by \cite{Wands:2000dp}
\begin{equation}
\zeta = \frac{\frac43\rho_r\zeta_r + \rho_c\zeta_c}{\frac43\rho_r + \rho_c} ,
\end{equation}
where $\zeta_i \equiv -\Phi + \delta_i/(3(1+w_i))$ is the curvature perturbation of the fluid $i$ on the uniform energy density hypersurface of fluid $i$ and $\Phi$ is the gravitational potential in the longitudinal gauge. The isocurvature perturbation can then be expressed as $S_{rc}=3(\zeta_r - \zeta_c)$. At early times $\zeta\simeq \zeta_r$, whereas at the time of photon decoupling
\begin{equation}
\label{zeta_final}
\zeta \simeq \zeta_r + \frac{z_{\rm eq}/z_{\rm dec}}{4+3z_{\rm eq}/z_{\rm dec}}|S_{rc}| ,
\end{equation}
where $z_{\rm eq}$ is the redshift to the matter-radiation equality and $z_{\rm dec}$ is the redshift to the CMB. Thus, in the presence of DM isocurvature, the final curvature perturbation is larger than in scenarios with purely adiabatic ($S_{rc}=0$) perturbations. The more blue-tilted the DM energy density spectrum is, the more small scale structure can be expected to form. However, as perturbations at different scales start evolving upon their horizon entry, in practice the curvature perturbation spectrum can be expected to peak at some intermediate scale whose exact location depends on details of the underlying scenario. This was recently studied in the context of non-minimally coupled scalar field DM in Ref. \cite{Alonso-Alvarez:2018tus}; see also Ref. \cite{Graham:2015rva} for the case of vector DM. 

Thus, we conclude that enhancement in small scale structure formation is a generic prediction of models where DM resides in a decoupled sector which also contains scalar fields. As the result \eqref{zeta_final} is independent of the origin of the DM perturbation spectrum (e.g. misalignment, large mass, DM self-interaction, coupling to the inflaton or gravity) our work generalizes the earlier findings in the literature and shows that all such cases predict characteristic enhancement in small scale structure formation. This is likely to allow one to test models where DM interacts with matter feebly or only gravitationally. Also, it is clear that the present model cannot accommodate sizable DM self-interactions which may be observable in the future \cite{Tulin:2017ara}. Thus, a detection would rule out the simplest scenario considered in this Letter and provide for a way to distinguish it from models which predict similar structure growth but exhibit sizable self-interactions.


In this Letter we have considered the simplest possible DM Lagrangian and evaluated the criteria for the scalar field $\chi$ to constitute or source all or part of the observed DM. We considered two scenarios: the usual case where the field dynamics was dominated by slow-roll during inflation (the misalignment mechanism), and for the first time a scenario where the field attained equilibrium between the classical drift and stochastic quantum fluctuations. As in all such cases the DM constitutes a genuine isocurvature component (without, however, being in conflict with the CMB data), we found that this class of scenarios generically predicts enhancement in formation of structures at different scales. We plan to investigate the consequences of this in more detail in forthcoming publications.


\textit{Acknowledgements}:
I thank C. Cosme, G. Dom\`{e}nech, M. Kamionkowski, T. Markkanen, D. Mulryne, V. Poulin, A. Rajantie, R. Schaefer, T. L. Smith, and J. V\"aliviita for enlightening discussions and acknowledge the Simons Foundation for funding.

\bibliography{Simplest_DM}

\end{document}